\documentclass[aps,prd,
nofootinbib,
superscriptaddress,preprintnumbers,twocolumn]{revtex4-1}

\usepackage{hyperref}
\usepackage{color}
\usepackage{graphicx}
\usepackage{amsmath,amssymb,amsfonts}
\usepackage{bm}
\usepackage{hyperref}
\usepackage{ulem}
\usepackage{lineno}
\usepackage[dvipsnames]{xcolor}

\def \be {\begin{equation}}
\def \ee {\end{equation}}
\def \bes {\begin{subequations}}
\def \ees {\end{subequations}}

\newcommand{\beq}{\begin{eqnarray}}
\newcommand{\eeq}{\end{eqnarray}}

\def \pd {\partial}
\newcommand{\sP}{\mathcal{P}}

\def \<{\langle}
\def \>{\rangle}
\def \+{\dagger}
\def \({\left(}
\def \){\right)}

\def \le{\left}
\def \ri {\right}

\def \[{\left[}
\def \]{\right]}
\def \a {\alpha}
\def \b {\beta}

\def \e {\varepsilon}
\def \g {\gamma}

\def \o {\omega}

\def \m {\mu}
\def \n {\nu}
\def \s {\sigma}

\def \O {\Omega}

\def \vp {\vec{p}}
\def \vq {\vec{q}} 
\def \vx  {\vec{x}}
\def \vy {\vec{y}}

\def \vv {\bm{v}}

\def \vgamma {\bm{\gamma}}

\def \Tr {{\rm Tr}}
\begin{document}


\title{
Spin Hall effect in heavy ion collisions
}

\author{Shuai Y.F.~Liu}
\affiliation{Quark Matter Research Center, Institute of Modern Physics, 
Chinese Academy of Sciences,
 Lanzhou, Gansu, 73000, China }

\author{Yi Yin}

\affiliation{Quark Matter Research Center, Institute of Modern Physics, 
Chinese Academy of Sciences,
 Lanzhou, Gansu, 73000, China }

 \affiliation{
 University of Chinese Academy of Sciences,
 Beijing, 100049, China
 }

\date{\today}

\begin{abstract}
Spin Hall effect (SHE) is the generation of spin current due to an electric field, and has been observed in a variety of materials. 
In this work, we use linear response theory to verify that the analogous spin Hall current can be induced by chemical potential and temperature gradient, 
both of which are present in hot and dense nuclear matter created in heavy-ion collisions. 
We propose to measure ``directed spin flow'', the first Fourier coefficients of local spin polarization of  $\Lambda$ ($\bar{\Lambda}$) hyperon, at central collisions to probe spin Hall current in heavy-ion collision experiments.
We benchmark the magnitude of the induced ``directed spin flow'' at two representatively collisions energies, namely $\sqrt{s}_{NN}=200~{\rm GeV}$ and $\sqrt{s}_{NN}=19.6~{\rm GeV}$, by employing a phenomenologically motivated freeze-out prescription.
At both beam energies, the resulting ``directed spin flow'' ranges from $10^{-4}$ to $10^{-3}$, 
and is very sensitive to the rapidity. 
\end{abstract}

\maketitle

\section{Introduction}
The study of spin current, the flow of spin, 
has triggered intense research. 
The generation of spin current is a key concept in the field of spintronics~\cite{spin-tronics-nature}, 
and can be employed to probe intriguing properties of quantum materials~\cite{SpinProbe}. 
One prominent mechanism of the generation of spin current is spin Hall effect (SHE)~\cite{RevModPhys.87.1213}, 
by which an electric field will induce a transverse spin current perpendicular to the direction of the electric field. 
SHE has been observed in a number of table-top experiments~\cite{PhysRevLett.95.226801,PhysRevLett.94.047204,RevModPhys.87.1213}.

When the temperature and/or density gradient is non-zero, 
they could induce analogous spin Hall current.
The resulting spin polarization distribution function of fermions (anti-fermion) $\vec{\sP}_{+}$ ($\vec{\sP}_{-}$), in the rest frame of the fluid, can be written as:
\begin{eqnarray}
\label{SHE}
\vec{\sP}^{{\rm SHE}}_{\pm}(\vp)=\s^{T}_{\pm}\, \frac{\vp}{\e_{\vp}} \times {\vec \pd} T+\s^{\mu}_{\pm}\, \frac{\vp}{\e_{\vp}} \times \(T{\vec \pd}(\frac{\mu}{T})\)\, ,
\end{eqnarray}
where $\vp$ denotes the spatial momentum. 
Here, $\s^{T,\mu}_{\pm}$ depends on temperature $T$ and chemical potential $\mu$ as well as the energy of fermion $\e_{\vp}=\sqrt{\vp^{2}+m^{2}}$ where $m$ is the fermion mass. 
In Eq.~\eqref{SHE}, 
thermodynamic force, collectively denoted by $\vec{F}={\vec \pd} T, T{\vec \pd} (\mu/T)$, plays the role of analogous electric field. 
We shall refer Eq.~\eqref{SHE} as the ``thermally-induced spin Hall effect'' (TSHE). 
The first term in Eq.~\eqref{SHE}, i.e. spin Hall current induced by temperature gradient, is also known as the spin Nernst effect (SNE), 
and has been observed in platinum~\cite{SNE-nature} and in W/CoFeB/MgO heterostructures~\cite{Shenge1701503} a few years ago.

In this work, 
we shall study TSHE~\eqref{SHE} in hot and dense nuclear matter created in heavy-ion collisions.
The spin polarization induced by vorticity $\vec{\o}$ and magnetic field in heavy-ion collisions has attracted much experimental~\cite{STAR:2017ckg,Niida:2018hfw,Adam:2018ivw,Adam:2019srw,Acharya:2019ryw,Acharya:2019vpe,Zhou:2019lun} and theoretical efforts~\cite{Gao:2012ix,Becattini:2013fla,Fang:2016vpj,Pang:2016igs,Becattini:2017gcx,Florkowski:2019qdp,Liu:2019krs,Weickgenannt:2020aaf,Fu:2021pok}~(see Refs.~\cite{Kharzeev:2015znc,Bzdak:2019pkr,Becattini:2020ngo,Gao:2020vbh} for reviews). 
TSHE is different from the generation of spin polarization induced by fluid vorticity $\vec{\o}$.
To see such difference, let us assume for the sake of discussion, that $\s^{T,\mu}$ only depends on $p=|\vp|$.
We then have from Eq.~\eqref{SHE}, $\int_{\vp}\sP^{{\rm SHE}}=0$, meaning TSHE does not generate net polarization after the average over the momentum. 
In Ref.~\cite{1971JETPL..13..467D} where the notion of spin current is originally introduced, 
spin current is described by a tensor ${\cal S}^{ij}$. 
The first index of ${\cal S}^{ij}$ indicates  the  direction  of  flow,  while  the  second one indicates which component of the spin is flowing, i.e.,
 ${\cal S}^{ij}\propto\int_{\vp}p^{i} \sP^{j} $.
 Eq.~\eqref{SHE} then implies $S^{ij}\propto \epsilon^{ijk}\,F_{k}$, the generation of spin current. 
In contrast, in the same setting, turning on a vorticity will simply induce a net polarization $\propto \vec{\o}$, but will not induce spin current $S^{ij}$.
Therefore, the study of TSHE complements the existing work on the vorticity effects.

This work is organized as follows.
To quantify induced spin Hall current, we determine $\s^{T}_{\pm}, \s^{\mu}_{\pm}$ by evaluating the relevant correlation functions based on the linear response theory in Sec.~\ref{sec:theory}, 
see Eqs.~\eqref{sH}, \eqref{sT} for explicitly expression.
In Sec.~\ref{sec:HIC}, 
we then estimate the magnitude of observables associated with local spin polarization due to TSHE  at two representative beam energies, namely $\sqrt{s}_{NN}=200$~GeV and $\sqrt{s}_{NN}=19.6$~GeV,  
and will focus on central collisions.
It is well-known that the effects of temperature gradient and fluid vorticity in combination as the ``thermal vorticity'' will induce a specific pattern in the azimuthal angle dependence of $P^{i}(\phi)$ non-central collisions~\cite{Pang:2016igs,Becattini:2007nd,Becattini:2007sr,Becattini:2013fla,Wu:2019eyi} (see Ref.~\cite{Becattini:2020sww} for a review).
However, the analysis of local spin polarization in central collisions has attracted much less attention. 
We summarize our results in Sec.~\ref{sec:conclusion}

%

\begin{figure}
\centering
\includegraphics[width=0.25\textwidth]{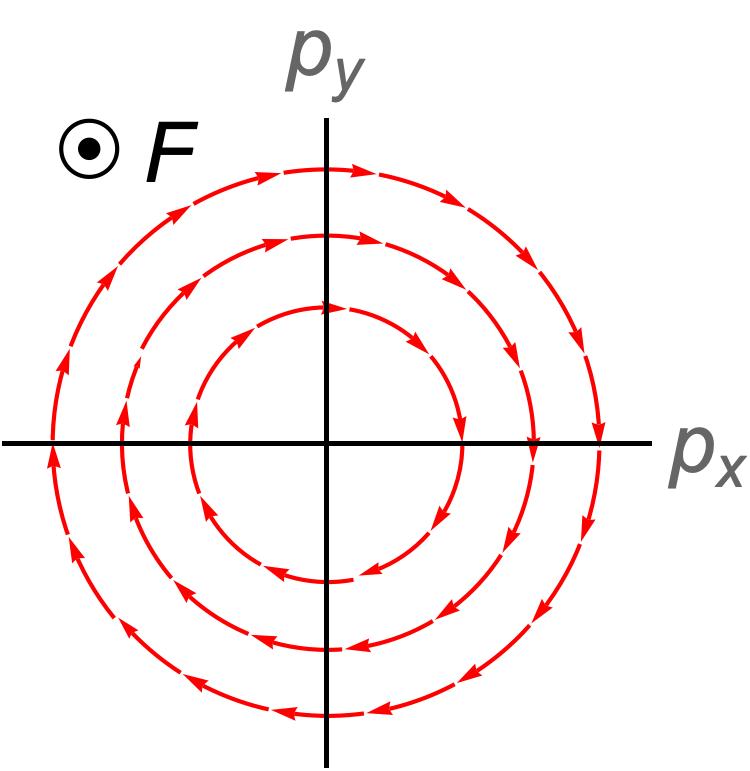}
\caption{
\label{fig:demo}
(color online)
A sketch illustrating the spin polarization (of particle) in momentum space induced by a thermodynamic force ${\vec F}={\vec \pd}T, T{\vec \pd}(\mu/T)$ to Eq.~\eqref{SHE}. 
Here, we choose $x-y$ plane to be the plane transverse to the direction of ${\vec F}$, and $p_{x,y}$ denote the momentum of fermions projected to $x,y$-directions.  
The arrows show the direction of spin polarization. 
 }
\end{figure}

\section{Thermal field theory calculations
\label{sec:theory}
}

In this section, 
we explain in detail on how to determine $\s^{T,\mu}_{\pm}$ which quantify TSHE by evaluating the relevant correlation functions.
The method used here is similar to the one developed by Luttinger in Ref.~\cite{PhysRev.135.A1505} to derive Kubo formula for thermal conductivity. 
Below, we consider the simplest case when there is only one species of fermions (anti-fermions) with one unit of $U(1)$ charge throughout, although relaxing this simplification is straightforward. 
We have also assumed that fermionic constituents of the systems are weakly coupled so that we can obtain those correlation functions from an one-loop thermal field calculation.

Let us begin with the operator:
\begin{eqnarray}
\label{sP-operator}
\hat{{\sP}}^{i}(t,\vx,\vy)\equiv \bar{\psi}(t,\vx+\frac{\vy}{2}) \gamma^{5}\g^{i}\, \psi(t,\vx-\frac{\vy}{2})\, ,
\end{eqnarray}
where $\g^{5},\g^{i}$ denote the standard gamma matrices. 
At this point,  $\psi$ represents a generic Dirac field. 
According to the quantum field theory, 
the phase space distribution of spin polarization is given by the Wigner transform:
\begin{eqnarray}
\label{sP-def}
\sP^{i}(t,\vx;\vp)=\int d^{3}\vy\langle \hat{{\sP}}^{i}(t,\vx,\vy) \rangle\,e^{i \vp\cdot \vy}\,
\end{eqnarray}
with $\langle\ldots\rangle$ denoting the thermal ensemble average. 
Here $\vec{\sP}$ include contribution from both particles and anti-particles.

We shall consider a medium that is initially in equilibrium and is isotropic, and hence at this point, $\vec{\sP}=0$.  
We next turn on an electric field and a non-flat metric 
\begin{eqnarray}
g^{\mu\nu}=(-1+2\phi(t,\vx),1,1,1)
\end{eqnarray}
where $\eta_{\m\n}$ is the flat-space metric.
Such external fields will in turn induce temperature and chemical potential gradient. 
${\vec \sP}$ arises due to the nonuniformity of the system and should be expressible in terms of a gradient expansion as
\begin{align}
\label{SHE-with-E}
\vec{\sP}_{\pm}=\frac{\vp}{\e_{\vp}} \times\le[ 
\s^{T}_{\pm}\, \vec{\pd} T+\s^{\phi}_{\pm}\,T{\vec\pd}\phi+\s^{\mu}_{\pm}\, (T{\vec \pd}(\frac{\mu}{T}))
+\s^{E}_{\pm}\, \vec{E}
\ri]\, ,
\end{align}
where $\s^{T,\phi,\mu,E}_{\pm}$ would depend on $T,\mu$ and $p$. 
Here, it is not necessary to include the acceleration of the velocity field $\pd_{t}\vv$ which can be expressed as a specific linear combination of gradient terms shown in Eq.~\eqref{SHE-with-E} using the ideal hydrodynamic equations.

Following Ref.~\cite{PhysRev.135.A1505}, we then consider the behavior of constitutive relation Eq.~\eqref{SHE-with-E} in two different limits:
a) The `` rapid'' case, where the typical frequency is much larger than gradient b) the ``slow'' case, in which the typical gradient is much larger than frequency. 
In the ``rapid'' case, the system does not have time to adjust to generate temperature and chemical potential gradient even if the external fields are present, so we have 
\begin{align}
\label{rapid}
   \lim_{r}  \vec{\sP}_{\pm}
   = \s^{\phi}_{\pm}\,T{\vec\pd}\phi+\s^{E}_{\pm}\, \vec{E}\, .
\end{align}
In the opposite limit, 
the system will reach hydrostatic state, in which $\vec{E}=T\nabla(\mu/T)$ and $T^{-1}{\vec \pd} T={\vec \pd}\phi$, and hence we have:
\begin{eqnarray}
\label{slow}
\lim_{s}\vec{\sP}_{\pm}= \,\le[ \le( \sigma^{T}_{\pm}+\s^{\phi}_{\pm}\ri)\,  T\vec{\pd}\phi + (\s^{\mu}_{\pm}+\s^{E}_{\pm})\vec{E}\ri]\, . 
\end{eqnarray}
Taking the difference between \eqref{slow} and \eqref{rapid}, 
we obtain
\begin{align}
\label{r-s}
\lim_{s}\, \vec{\sP}_{\pm}-\lim_{r}\, \vec{\sP}_{\pm}
= \s^{T}_{\pm}\,T{\vec\pd}\phi+\s^{\mu}_{\pm}\, \vec{E}
\end{align}

Based on Eq.~\eqref{r-s}, we can extract $\s^{T,\mu}_{\pm}$ from the relevant retarded correlation functions:
\begin{eqnarray}
\label{GE-def}
  &\,&  G^{i}(t,\vx;\vy)=i\, \langle \hat{\sP}^{i}(t,\vx;\vy)\, \hat{J}^{0}(0,0;0) \rangle \theta(t)\, ,
    \\
\label{R-def}
  &\,&    G^{i,00}(t,\vx)= i\langle \sP^{i}(t,\vx)\, T^{00}(0,0) \rangle \theta(t)\, ,
\end{eqnarray}
where $\hat{J}^{0}=\bar{\psi}\g^{0}\psi$ and $\hat{T}^{\mu\nu}$ denotes the stress-energy tensor.
Introducing Fourier transform with $q_{0},\vq$ being frequency and momentum conjugate to $t,\vx$ respectively, we have
\begin{eqnarray}
\label{response}
\sP^{i}(q_{0},\vq;\vp)&=& 
 G^{i}(q_{0},\vq;\vp)\, A_{0}(q_{0},\vq)
\nonumber \\
&+&G^{i,00}(q_{0},\vq;\vp)\phi(q_{0},\vq)
\, .
\end{eqnarray}
In what follows, 
we shall first evaluate $G$ at one loop, and then extract $\s^{T,\mu}_{\pm}$ by comparing Eq.~\eqref{response} with Eqs.~\eqref{r-s}.

\begin{figure}[b]
\centering
\includegraphics[width=0.25\textwidth]{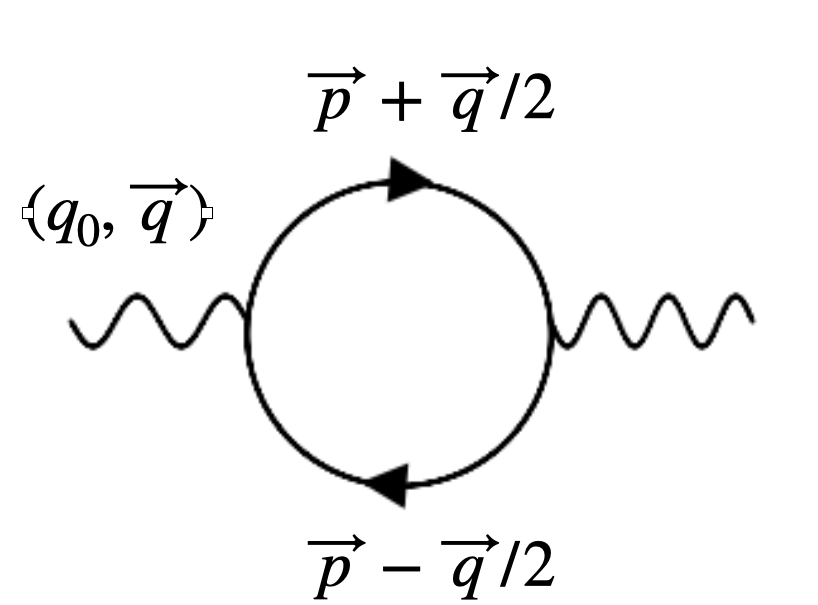}
\caption{
\label{fig:loop}
One loop diagram contributing to $G^{i}$ and $G^{i,00}$. 
}
\end{figure}

At one loop order (see Fig.~\ref{fig:loop}), 
 $G^{i}(\tilde{\o}_{n},\vq;\vp)$ as a function of the Bosonic Matsubara frequency $\tilde{\o}_{n}=2n\pi T$ is given by
\begin{align}
\label{Gi0}
G^{i}
=T\sum_{\nu_{m}}\Tr
\left[
\g^{i}\g^{5}
S\left(i \nu_{m}+i\tilde{\o}_{n},\vp_{1}\right)\g^{0}
S(i\nu_{m},\vp_{2})
\right],
\end{align}
where $\vp_{1}= \vp+\frac{\vq}{2},\vp_{2}= \vp-\frac{\vq}{2}$. 
Here the Euclidean propagator as function of the Fermionic Matsubara frequency $\nu_{n}=\pi T(2n+1)+\mu$ and momentum $\vp$ reads 
\begin{eqnarray}
S(i\nu_{m},\vp)
= \sum_{s=\pm}\,\Lambda_{s}(\vp)\,  \Delta_{s}(i\nu_{m},\vp)\, , 
\end{eqnarray}
with $\Lambda_{s}(\vp)= s\g^{0}\e_{\vp}-\vp\cdot \vgamma +m$, 
and
\begin{eqnarray}
\Delta_{s}(i\nu_{m},\vp)= (\frac{-s}{2\e_{\vp}})\, \frac{1}{i \nu_{m}-s\e_{\vp}}\, . 
\end{eqnarray}

%

To evaluate Eq.~\eqref{Gi0},
we first take the trace:
\begin{eqnarray}
\label{Trace}
\Tr
\left[
\g^{i}\g^{5}
\Lambda_{s}(\vp_{1})\, \g^{0}\,\Lambda_{s'}(\vp_{2})
\right]
=4i\e^{ijm}\,q_{j}p_{m}\, ,
\end{eqnarray}
If one were computing the response of axial current $\vec{j}_{5}=\int d^{3}\vp/(2\pi)^{3}\vec{\sP}$ to external disturbance by loop diagrams, 
one has to integrate out momentum in the loop (e.g. Ref.~\cite{Lin:2018aon}).
However, since we are interested in $\vec{\sP}$, we only need to use the book-keeping formula to perform the summation over the Matsubara frequency:
\begin{eqnarray}
\label{M-sum}
&\,&T\sum_{\O_{m}} \Delta_{s}(i\O_{m}+i\tilde{\o}_{n},\vp_{1})\Delta_{s'}(i\O_{m},\vp_{2})
\nonumber \\
&=&
\sum_{ss'}\left(
\frac{-s s'}{4\e_{1}\e_{2}}
\right) \(\frac{s n_{s}(\e_{1})- s' n_{s'}(\e_{2})}{\tilde{\o}_{m}-s\e_{1}+s'\e_{2}}\)
\end{eqnarray}
where $\e_{1,2}=\e_{\vp_{1,2}}$. 
Here, $n_{\pm}(\e)=1/(e^{(\e\mp\mu)/T}+1)$ are Fermi-Dirac distribution for fermion($+$) and anti-fermion ($-$). 
By substituting Eq.~\eqref{Trace} and Eq.~\eqref{M-sum} into Eq.~\eqref{Gi0}, and perform the analytic continuation, which
amounts to replace $\tilde{\o}_{m}$ with $q_{0}+i0^{+}$, 
we obtain the desired expression
\begin{eqnarray}
\label{Gi0-1}
&\,&
G^{i}(q_{0},\vq;\vp)
=-\e^{iml}\frac{p^{m}}{\e^{2}_{\vp}}\, iq^{l}\,
\Big\{
\frac{\vq\cdot\vv_{\vp}}{q_{0}-\vq\cdot\vv_{\vp}+i 0^{+}}\frac{\pd n_{+}(\e_{\vp})}{\pd \e_{\vp}}
\nonumber \\
&-&\frac{\vq\cdot\vv_{\vp}}{q_{0}+\vq\cdot\vv_{\vp}+i 0^{+}}\frac{\pd n_{-}(\e_{\vp})}{\pd \e_{\vp}}
+ \frac{n_{+}(\e_{\vp})+n_{-}(\e_{\vp})}{\e_{\vp}}
\Big\}
\end{eqnarray}
Here,we have used $\e_{1}-\e_{2}=\vv_{\vp}\cdot\vq, n_{\pm}(\e_{1})-n_{\pm}(\e_{2})=(\pd n_{\pm}(\e_{\vp})/\pd \e_{\vp})\vv_{\vp}\cdot\vq$
by assuming $q_{0},q\ll \e_{\vp}\sim T, \mu$. 
In another word, we have expanded $G^{i}$ to the first non-trivial order in $q_{0}/\e_{\vp},q/\e_{\vp}$ to obtain \eqref{Gi0-1}.

At one loop order
the expression for $G^{i,0 0}$ is given by replacing
$\g^{0}$ in Eq.~\eqref{Gi0} with $\g^{0}i\nu_{m}$.
Hence the evaluation of $G^{i,00}$ becomes a trivial extension of the calculation of $G^{i}$, yield
\begin{eqnarray}
\label{Gi00-1}
G^{i,00}(q_{0},\vq;\vp)
&=&- i\e^{ilm}p^{l}\frac{q^{m}}{\e_{\vp}}\,
\Big\{
\frac{\vq\cdot\vv_{\vp}}{q_{0}-\vq\cdot\vv_{\vp}+i 0^{+}}\frac{\pd n_{+}(\e_{\vp})}{\pd \e_{\vp}}
\nonumber \\
&+&
\frac{\vq\cdot\vv_{\vp}}{q_{0}+\vq\cdot\vv_{\vp}+i 0^{+}}\frac{\pd n_{-}(\e_{\vp})}{\pd \e_{\vp}}
\Big\}\, , 
\end{eqnarray}

We now extract the induced fermion (anti-fermion) local spin polarization $\vec{\sP}_{+}$ ($\vec{\sP}_{-}$) by substituting Eq.~\eqref{Gi0-1} Eq.~\eqref{Gi00-1} into Eq.~\eqref{response}. 
We assume that $\sP_{+}$ ($\sP_{-}$) should depend on $n_{+}$ ($n_{-}$) and/or the derivative $\pd n_{+}/\pd \e_{\vp}$ ($\pd n_{-}/\pd \e_{\vp}$) only, 
and require $\vec{\sP}(\vp)=\vec{\sP}_{+}(\vp)+\vec{\sP}_{-}(-\vp)$. 
Use the prescription outlined above, we obtain
\begin{eqnarray}
\label{P-and-g}
\sP_{\pm}&=& \frac{\vp}{\e_{\vp}}\times
\Big [ g_{\pm,\mu}(q_{0},\vq;\mu)\, \vec{E}(q_{0},\vq)
\nonumber \\
&\,&+g_{\pm,T}(q_{0},\vq;p) T(i \vec{q} \phi)
\Big ]\, ,
\end{eqnarray}
where we have used $\vec{E}=-i \vq A_{0}$ and where
\begin{eqnarray}
\label{g-fun}
&\,&g_{\pm,\mu}=
\frac{\pm 1}{\e_{\vp}}\left[
\frac{\vq\cdot\vv_{\vp}}{q_{0}\mp\vq\cdot\vv_{\vp}+i 0^{+}}\frac{\pd n_{\pm}(\e_{\vp})}{\pd \e_{\vp}}
+ \frac{n_{\pm}(\e_{\vp})}{\e_{\vp}}
\right]\, .
\\
\label{Gi00-1}
&\,&g_{\pm,T}
= \frac{1}{T}
\frac{\vq\cdot\vv_{\vp}}{q_{0}\mp\vq\cdot\vv_{\vp}+i 0^{+}}\le(-\frac{\pd n_{\pm}(\e_{\vp})}{\pd \e_{\vp}}\ri)\, . 
\end{eqnarray}

The cautious reader might worry if one could do such separation when fermion and anti-fermion are mixed with each in the presence of an external field. 
However, since we consider fermions with energy $\e_{\vp}\gg |\vq|,q_{0}$, 
they will interact with anti-fermions of approximately the same energy $\e_{\vp}$, meaning the relative phase between the fermions and anti-fermions participated in such interaction is approximately $2\e_{\vp} \Delta t$ for a given duration $\Delta t$. 
Therefore one should be able to integrate out fast oscillating anti-fermions in the long time limit and hence obtain distribution for fermions. 
One can draw a parallel conclusion for anti-fermions.
See Refs.~\cite{Son:2012zy,Manuel:2014dza,Manuel:2016wqs} for explicit examples on obtaining particle/anti-particle distribution through such integrating-out procedure.

We now return to the extraction of $\s^{T,\mu}_{\pm}$. 
Comparing Eq.~\eqref{SHE-with-E} in Fourier space with Eqs.~\eqref{r-s} 
and using $\lim_{r}=\lim_{q_{0}\to 0}\lim_{\vq\to 0}$ and $\lim_{s}=\lim_{\vq\to 0}\lim_{q_{0}\to 0}$,
we find
\begin{eqnarray}
\label{sH}
\s^{\mu}_{\pm}&=&\lim_{s} g_{\mu}(q_{0},\vq)-\lim_{r} g_{\mu}(q_{0},\vq)
\nonumber 
\\
&=&\mp\frac{1}{\e_{\vp}}\left[
-\frac{\pd n_{\pm}(\e_{\vp})}{\pd \e_{\vp}}
\right] \, ,  
\\
\label{sT}
\s^{T}_{\pm}&=&\lim_{s} g^{T}_{\pm}(q_{0},\vq)-\lim_{r} g^{T}_{\pm}(q_{0},\vq)
\nonumber
\\
&=&\frac{-1}{T}\left[
-\frac{\pd n_{\pm}(\e_{\vp})}{\pd \e_{\vp}}
\right]\, .
\end{eqnarray}
Note $\s^{T}_{+},\s^{\mu}_{+}<0$.
In the massless limit, the expression of $\s^{\mu}$ in Eq.~\eqref{sH} agrees with that obtained in Ref.~\cite{Hidaka:2016yjf} using chiral kinetic theory.
This $\mu$-gradient induced spin polarization has also been obtained for a Landau fermion liquid with chiral fermions from quantum kinetic theory in Ref.~\cite{Son:2012zy} (see also Refs.~\cite{Hidaka:2016yjf,Hattori:2019ahi,Fang:2016vpj}. 
For related studies, see for example Refs.~\cite{Pu:2014fva,Sheng:2019kmk}). 
Our main results in this section, Eqs.~\eqref{sH}, \eqref{sT} apply to fermions with arbitrary mass in a system with generic $\mu, T$.

The authors of Refs.~\cite{Becattini:2007nd,Becattini:2007sr,Becattini:2013fla} consider a class of hydrodynamic profile which satisfies the so-called "equilibrium/static" condition~\cite{Becattini:2012tc,Becattini:2018duy}, i.e.,
\begin{align}
	\label{Killing}
	\pd_{\mu}\,(\beta u_{\nu})+\pd_{\nu}\,(\beta u_{\mu})=0
	\qquad
	\pd_{\nu}(\beta\,\mu)=0\, ,
\end{align}
and find that the spin polarization is proportional to the thermal vorticity
\begin{align}
\label{T-vorticity}
	\omega^{\rm th}_{\mu\nu}\equiv - \frac{1}{2}\, \left[\pd_{\mu}(\beta u_{\nu})-\pd_{\nu}(\beta u_{\mu})\right]\, . 
\end{align}
The thermal vorticity effects as defined in Eq.~\eqref{T-vorticity} include the contribution from both fluid vorticity (the anti-symmetric part of the flow velocity gradient) and temperature gradient. 
Note, the thermal vorticity also contains a term proportional to the acceleration $u^{\mu}\pd_{\mu} u^{\nu}$, but this term can be replaced by temperature gradient if one combines ideal hydrodynamic equation with the condition~\eqref{Killing}.
However, 
the effect of $\mu/T$ gradient is different from that of thermal vorticity since the condtion~\eqref{Killing} requires the homogeneity of $\mu/T$, meaning the analysis assuming such condition does not account for the contribution from $\mu/T$ gradient.

It is worth pointing out that the effects of different hydrodynamic gradients on spin polarization have to be extracted from different correlation functions. 
As we have just derived in this Section, 
the effects of $\mu$ and temperature gradient can be extracted from correlator $G^{i}$~\eqref{Gi0-1} and $G^{i,00}$~\eqref{Gi00-1} respectively. 
In Ref.~\cite{Liu:2021uhn}, 
we further show how to determine the fluid vorticity effect from $G^{i,0j}\equiv i\langle \sP^{i}(t,\vx)\, T^{0j}(0,0) \rangle \theta(t)$
\footnote{
As shown in later works.~\cite{Liu:2021uhn,Fu:2021pok,Becattini:2021suc}, shear tensor (the symmetric part of flow gradient) could also induce spin polarization in a specific way.
}. 
Since the physical content of those correlators are not identical,
we expect that those effects may generally tell us complimentary information on the spin physics of a medium.

\section{Phenomenology
\label{sec:HIC}
}

\subsection{Observables}

We now discuss the proposal for detecting spin Hall current in heavy-ion collision experiments. 
Our proposal relies on two elements. 
The first is made by noting in fireball created in those collisions, both temperature and baryon chemical potential gradient can be sizable. 
The second element is that  
the average differential spin polarization vector of $\Lambda$ and $\bar{\Lambda}$, $P^{i}_{\pm}(\phi_{p})$, as a function of azimuthal angle $\phi_{p}$ are measured experimentally to good precision through the angular distribution of the decay daughters of $\Lambda,\bar{\Lambda}$~\cite{Adam:2019srw}.
According to Eq.~\eqref{SHE},
the induced local spin polarization $\sP$ projected into the transverse plane will feature a dipole pattern, see Fig.~\ref{fig:demo}. 
We then propose to use the first Fourier coefficients of $P^{i}_{\pm}(\phi_{p})$ ($i=x,y$) to probe the resulting spin current (see also Ref.~\cite{Xia:2018tes}): 
\begin{eqnarray}
\label{av-def}
\left( a^{i}_{1,\pm},
v^{i}_{1,\pm}
\right)
\equiv
\int \frac{d\phi_{p}}{2\pi} P^{i}_{\pm}\,\times \left(\sin\phi_{p},\cos\phi_{p} \right)\, . 
\end{eqnarray}
The first Fourier harmonics of produced hadrons in heavy-ion collisions, i.e., ``directed flow'', are employed to measure the flow of those hadrons.
Motivated by this, we will refer  $a^{i}_{1,\pm},v^{i}_{1,\pm}$ as \textit{``directed spin flow''}.

%
%
%
As a premier, 
we consider following freeze-out prescription connecting $P^{i}$ and $\sP^{i}$: 
\begin{align}
\label{P-freezeout}
P^{i}_{\pm}(\phi_{p})=\frac{\int_{y}\int_{p_{T}}\int  d\Sigma^\alpha\,p_\alpha \mathcal{P}^{i}_{\pm,{\rm lab}}}{2\int_{y}\int_{p_{T}}\int  d\Sigma^\a\,p_\a \, n_{\pm}(\e')}
\end{align} 
where $\Sigma^{\mu}$ denotes the freezeout hyper-surface and the factor of $2$ in the denominator accounts for $2$ spin states of $\Lambda$ 
($\bar{\Lambda}$). 
A similar prescription has been used to study spin polarization induced by ``thermal vorticity''~\cite{Becattini:2013fla}. 
Here, 
we boost the result in the fluid rest frame to the lab frame.
For example, for effects induced by ${\vec \pd} T$ (the first term in Eq.~\eqref{SHE}), 
we use
\begin{eqnarray}
\label{Pmu-general}
\sP^{\rho,\mu}_{\pm,{\rm lab}}=\frac{1}{\e'}
\e^{\rho\nu\a\b}u_{\nu} p_{\a}\,
\s^{T}_{\pm}(\e',\mu_{B})\pd_{\b}T\, , 
\end{eqnarray}
where $\e'=p^{\mu}\cdot u_{\mu}$ and $u^{\mu}$ is the four flow velocity.
The treatment of the second term in Eq.~\eqref{SHE} is similar.
In Eq.~\eqref{P-freezeout},
the integration over transverse momentum $p_{T}$ and momentum rapidity $y$ reads
\begin{eqnarray}
\label{y-range}
\int_{p_{T}}\equiv \int^{p_{T,{\rm max}}}_{0} \frac{dp_{T}}{2\pi} p_{T}\, ,
\qquad
\int_{y}\equiv \int^{y_{c}+\Delta y}_{y_{c}-\Delta y} \,dy\, .
\end{eqnarray}

Since $\mu_{B}$ and $T$ is symmetric in the spatial rapidity $\eta_{s}$ on the freezeout surface,
following Ref.~\cite{Brewer:2018abr}, 
we assume that the deviation from boost invariance takes the form
\begin{eqnarray}
\label{mu-param}
\label{Tmu-param}
T(\eta_{s})= T_{f}-\a_{T}\, \eta^{2}_{s}\, , 
\qquad
\mu_{B}(\eta_{s})=\mu_{B,f}+\a_{\mu} \eta^{2}_{s}\, ,
\end{eqnarray}
with $T_{f},\mu_{B,0}$ and $\a_{T}, \a_{\mu}$ depend on the beam energy $\sqrt{s}_{NN}$. 
We shall use this form for illustrative purposes,
noting of course that it cannot be relied upon at large $\eta_{s}$.

While we use Eq.~\eqref{mu-param} to compute $\mu_{B}$ gradient in Eq.~\eqref{Pmu-general},
we shall evaluate $\s^{T,\mu}_{\pm}(\e',\mu), n_{\pm}(\e')$ in Eq.~\eqref{Pmu-general} using the parametrization of the flow profile on the freezeout surface based on a blastwave model.
Since we are considering central collisions, we will use flow profile which is boost-invariant and spherically symmetric, 
and further assume the freeze-out surface is isochronous at $\tau_{f}$. 
This treatment is consistent with our formalism based on the linear response as adding non-boost invariant corrections to the evaluation of $\s^{T,\mu}, n_{\pm}(\e'), \Sigma^{\mu}$ would lead to contribution at higher-order in gradient.

\begin{table}[h]
	\begin{tabular}{ | l |c|c|c|c| r |}
		\hline
		$ \sqrt{s}_{NN} $&$ T_{f} $  &$\alpha_T$& $\mu_{B,f}$&$\alpha_\mu$ & $ v\qquad $\\ \hline
		$200$~GeV&89~MeV & 0.033$ T_f $& 22~MeV&11~MeV & $ 0.835(r/R)^{0.82} $\\ \hline
		$19.6$~GeV&113~MeV & 0.075$ T_f $& 196~MeV&50~MeV&$ 0.664(r/R)^{0.90} $\\ 
		\hline
	\end{tabular}
	\caption{The parameters for $ \sqrt{s}=200$~GeV(first row) and $ \sqrt{s}=19.6$~GeV (second row).
	}
	\label{tab_para}
\end{table}

Using the setup specified above, we perform the calculations with the parameters listed in Table.~\ref{tab_para}.
Here,
the kinetic freezeout temperature ($T_f $), transverse flow velocity ($v$) and $ \mu_{B,f} $ are taken from those in Refs.~\cite{Abelev:2008ab,Adamczyk:2017iwn}. 
Our choice of the values of $ \alpha_\mu $ is guided by Ref.~\cite{Becattini:2007ci}. 
To estimate $\a_{T}$,
we consider the pion rapidity distribution in  $dN/dy$, which are measured and fitted with a Gaussian shape $dN/dy\sim \exp(-y^{2}_{s}/(2\s^{2}))$ in Ref.~\cite{Bearden:2004yx,Flores:2016mtp}. 
We will relate the experimentally extracted width $\s(\sqrt{s}_{NN})$ with $\a_{T}$ by assuming $dN/d y$ as a function of momentum rapidity $y$ as a proxy for the spatial rapidity distributions of entropy. 
Further assuming the produced entropy scales as $T^{3}$, 
we have $T(\eta_{s}) \sim \exp(-\eta^{2}_{s}/(6\s^{2}))$, and obtain $\a_{T}=T_{f}/(6\s^{2}) $ by matching the small $\eta_{s}$ behavior of $T(\eta_{s})$ to Eq.~\eqref{Tmu-param}.
Using $\s(200~{\rm GeV}) \approx 2.25$ ~\cite{Bearden:2004yx} and $\s(19.6~{\rm GeV}) \approx 1.5$~\cite{Flores:2016mtp},
we then have the values of $\a_{T}$ listed in Table.~\ref{tab_para}.
Finally, we benchmark the freezeout time with $\tau_{f}=10$~fm. 
We note our results will simply scale with $1/\tau_f$.

\begin{figure*}[t!]
  \includegraphics[width=1\textwidth]{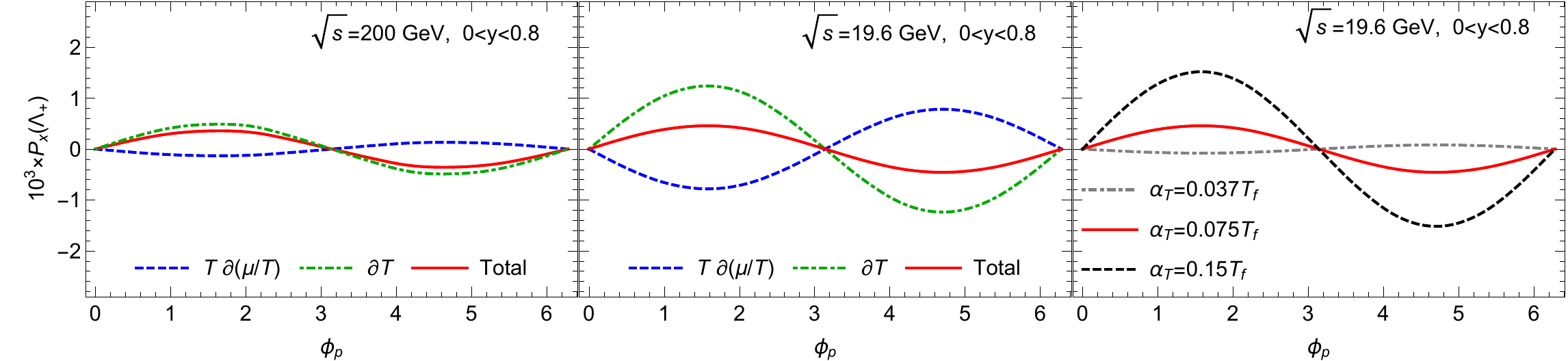}%
\caption{
\label{fig:pol}
(color online)
We show the signature of spin Hall current induced by baryon chemical potential gradient and temperature gradient by plotting the $x$-component of the polarization of $\Lambda$, $P^{x}_{+}$, as a function of azimuthal angle $\phi_{p}$ at $\sqrt{s}_{NN}=200$~GeV (left) and $19.6$~GeV (middle) for central collisions using Eqs.~\eqref{SHE}, \eqref{sH}, \eqref{P-freezeout} and a blastwave model. 
In the Figure, the green dashed and blue dashed curves show the contribution from ${\vec \pd}T$ and $T {\vec \pd}(\mu/T)$, respectively. 
To illustrate the sensitivity of SHE signature to the temperature and chemical potential profile, 
we consider $P^{x}_{+}$ at $\sqrt{s}_{NN}=19.6$~GeV using three different values of $\a_{T}$ (see Eq.~\eqref{Tmu-param}).
In additional to $\a_{T}=0.075~T_{f}$ used in obtaining the middle figure, we consider  $\a_{T}=0.15, 0.0375~T_{f}$. 
}
\end{figure*}

\subsection{Results}
In this Section, we demonstrate that azimuthal angle dependence of local spin polarization $\vec{P}$ in the transverse plane is sensitive to the spin Hall current induced by the longitudinal temperature and baryon chemical potential gradient in a way that yields distinctive, qualitative, observable consequences.  
Furthermore, we make an order of magnitude estimation for the resulting ``directed spin flow'' observables.

In Fig.~\ref{fig:pol}, we compute azimuthal angle $\phi_{p}$ dependence of spin polarization vector of $\Lambda$ projected onto $x$-direction, $P^{x}_{+}(\phi_{p})$, at two representative beam energies, $\sqrt{s}_{NN}=200$~GeV and $19.6$~GeV.
Keeping in mind that an upgrade of the inner Time Projection Chamber (iTPC) at STAR will extend its rapidity acceptance for protons from
$|y|\leq 0.5$ to $|y|\leq 0.8$, 
we have integrated over the rapidity in the range $0<y<0.8$.
In addition, we use $p_{T,{\rm max}}=5$~GeV of the integration over $p_{T}$ in \eqref{y-range}.
At both $\sqrt{s}_{NN}$, $P^{x}_{+}(\phi_{p})$ exhibits the characteristic sinusoidal behavior with a period of $2\pi$, as we anticipate from Fig.~\ref{fig:demo}.
Because of the rotational symmetry in the transverse plane, 
$\vec{P}$ projected along any direction on the transverse plane $\hat{e}_{\perp}$, $P^{\perp}_{\pm}=\vec{P}_{\pm}\cdot \hat{e}_{\perp}$ can be simply related to $P_{x}(\phi_{p})$ by a phase factor, for example $P^{y}_{\pm}(\phi_{p})=P^{x}_{\pm}(\phi_{p}-\pi/2)$. 
We will focus on $P^{x}_{\pm}(\phi_{p})$ and $a^{x}_{1,\pm}$ from now on.

While $P^{x}_{+}(\phi_{p})$ is mainly induced by temperature gradient at $\sqrt{s}_{NN}=200$~GeV, 
$P^{x}_{+}(\phi_{p})$ that arises from $\vec{\pd} T$ and from $T\vec{\pd}(\mu/T)$ are comparable in magnitude, but oppose in sign,  at $\sqrt{s}_{NN}=19.6$~GeV (see the middle column of Fig.~\ref{fig:pol} ) within the current model set-up. 
Indeed, $P^{x}_{+}(\phi)$ is expected to be very sensitive to the details of temperature and chemical potential profile in heavy-ion collisions, and would depend on beam energy nontrivially. To further demonstrate this point, we show $P^{x}_{+}(\phi)$ by using two additional choices of $\a_{T}$ at $\sqrt{s}_{NN}=19.6$~GeV, i.e. $\a_{T}=0.037T_{f}$ and $\a_{T}=0.15T_{f}$. We observe even the sign of $P^{x}_{+}(\phi)$ would become different under different values of $\a_{T}$.
This observation in turn suggests that one might employ $P^{x}_{\pm}$ to probe temperature and chemical profile of the matter created in heavy-ion collisions.

\begin{figure}
  \includegraphics[width=.4\textwidth]{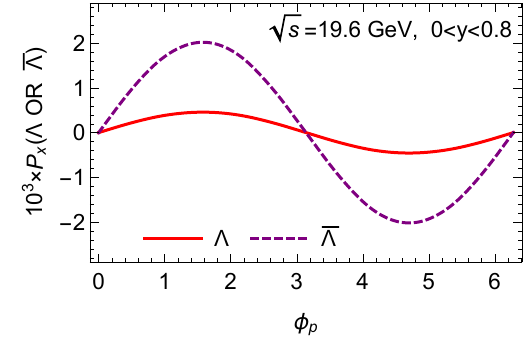}%
\caption{
\label{fig:polpm}
(color online)
We show the $x$-component of $\Lambda$ polarization $P^{x}_{+}$ and $\bar{\Lambda}$ polarziation $P^{x}_{-}$ induced by the temperature and  baryon chemical potential gradient for a central collision at $\sqrt{s}_{NN}=19.6$~GeV. 
Since local spin polarization induced by $\mu{\vec{\pd}}(\mu/T)$ is opposite in sign for $\Lambda$ and $\bar{\Lambda}$, we observe the splitting between $P^{x}_{+}$ and $P^{x}_{-}$. 
}
\end{figure}

Since ${\vec \pd}T$ and $T\vec{\pd}(\mu/T)$ are even and odd under charge-parity, 
the local spin current induced by the former is of the same sign for both $\Lambda$ and $\bar{\Lambda}$, 
but is of the opposite sign for that induced by the later (see also Eq.~\eqref{sH}). 
In Fig.~\ref{fig:polpm}, we
show $P^{x}_{+}(\phi_{p})$ and $P^{x}_{-}(\phi_{p})$ at $\sqrt{s}_{NN}=19.6$~GeV, 
and observe a sizable splitting between them. 
Therefore charge-dependent local spin polarization $P^{x}_{+}(\phi_{p})-P^{x}_{-}(\phi_{p})$ and charge-independent local spin polarization $P^{x}_{+}(\phi_{p})+P^{x}_{-}(\phi_{p})$ can help distinguish SHE induced by temperature and chemical potential gradient respectively.

\begin{figure}
  \includegraphics[width=.45\textwidth]{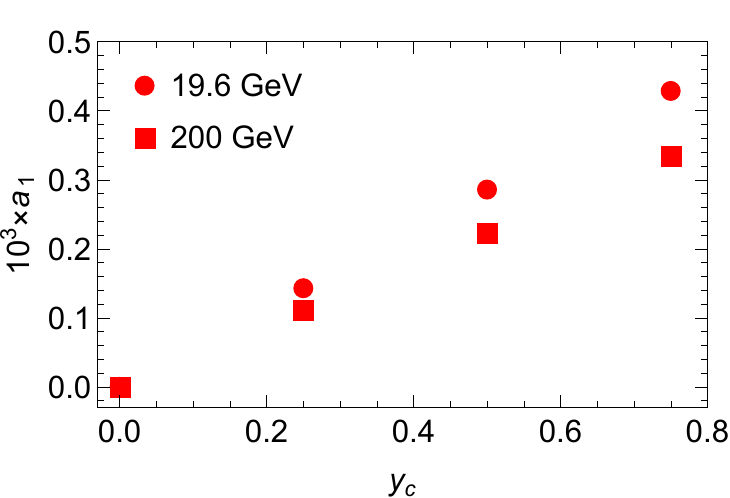}%
\caption{
\label{fig:a1}
We show ``directed spin flow'', $a^{x}_{1,\pm}$ (defined in Eq.~\eqref{av-def})  induced by temperature and baryon chemical potential gradient, at $\sqrt{s}=200,19.6$~GeV v.s. the center of the rapidity bin $y_{c}$ (see Eq.~\eqref{y-range}). 
}
\end{figure}

Next, we take a look at the magnitude of $a^{x}_{1,\pm}$ which quantifies spin Hall current. 
We can make a quick estimation as follows. 
We first  replace $-\pd n_{+}(\e_{\Lambda,p})/\pd \e$ with  $T^{-1}_{f} n_{+}(\e_{\Lambda})$ in Eq.~\eqref{sH} where $\e_{\Lambda}\sim 1$~GeV is the characteristic energy of $\Lambda$. 
From Eq.~\eqref{Tmu-param}, 
we may use  $|\vec{\pd} T| \sim \a_{T}\eta_{s}/\tau_{f}, |\vec{\pd} \mu| \sim \a_{\mu}\eta_{s}/\tau_{f}$.
We then have from Eqs.~\eqref{P-freezeout}, \eqref{Pmu-general} that $a^{x}_{1}$ induced by temperature gradient and chemical potential gradient is of the order $ (T_f/M)^{1/2}\a_{T}/(T^{2}_{f}\tau_{f})$, $ (T_f/M)^{1/2}\a_{\mu}/(\e_{\Lambda}T_{f}\tau_{f})$ respectively, where $M$ denotes the mass of $\Lambda$. 
With all the values of $\a_{T,\mu}$ and $T_{f},\tau_{f}$ in place, 
we find $a^{x}_{1}$ ranges from $10^{-4}$ to $10^{-3}$. 
This estimation is indeed consistent with the results shown in Fig.~\ref{fig:a1}. 
Although further work will be needed in order to check the present estimation quantitatively,
our results provide guidance on the feasibility of detecting SHE experimentally.

Motivated by the expanded rapidity coverage that the STAR iTPC upgrade will bring, and proposed forward rapidity program at both RHIC~\cite{YANG2019951} and LHC~\cite{Hadjidakis:2018ifr}, 
we finally consider the rapidity dependence of ``directed spin flow'' in Fig.~\ref{fig:a1}. 
We computed $a^{x}_{1,+}$ using Eq.~\eqref{y-range} with three different values of $y_{c}$, namely $y_{c}=0,0.25,0.5,0.75$ and with $\Delta y=0.25$ fixed. 
We observe $a^{x}_{1,+}$ increases with a larger $y_{c}$. 
This behavior arises from the parametrization of temperature and chemical potential profile we have used in Eq.~\eqref{Tmu-param} for the present illustrative purpose. 
However, since temperature and chemical potential would strongly depend on rapidity at large rapidity in heavy-ion collisions, 
we do expect that the signatures of ``directed spin flow'' would become more pronounced and become very sensitive to the rapidity bin at forward rapidity. 
Thus, the study of SHE will enrich the physics topics at planned forward rapidity program in heavy-ion collisions~\cite{YANG2019951,Hadjidakis:2018ifr}.

\section{Conclusions
\label{sec:conclusion}
}

In this work, we investigate the perspective of detecting spin Hall current induced by temperature and chemical potential gradient, i.e., thermally-induced spin Hall effect (TSHE),             in fireballs created in heavy-ion collisions. 
We use linear response theory to derive those effects. 
Complementary to the measurement of spin polarization induced by vorticity, 
the exploration of spin Hall current will provide a new probe to the quantum transport phenomenon of QCD matter.
We demonstrate that the induced spin Hall current would manifest itself through the azimuthal angle dependence of the local spin polarization of $\Lambda$ and $\bar{\Lambda}$, $P^{i}(\phi_{p})$, yielding a qualitative distinctive signature. 
We propose to use the first Fourier harmonics of $P^{i}(\phi_{p})$, ``directed spin flow'',  to quantify spin Hall current. 
We estimate the magnitude of ``directed spin flow'' is of the order $10^{-4}-10^{-3}$, 
and show it can be very sensitive to the rapidity.

We have made simplifications at many points, particularly on the parametrization of density and flow profile on the freeze-out surface, for illustrative purposes. 
We have limited ourselves to the discussion of spin current induced by longitudinal $\mu_{B}$ and/or $T$ gradient, 
but the presence of gradient in the transverse plane could lead to possible observable effects as well. 
Therefore future studies based on state of the art hydrodynamic modeling~\cite{Shen:2020gef} are desirable. 
As far as background contribution is concerned, 
we note when the spin polarization induced by some specific local vorticity profile is boosted by the bulk flow, it might contribute to ``directed spin flow'' observables, see for example Ref~\cite{Xia:2018tes}.
We point out however the direction of spin Hall current induced by the gradient of $\mu_{B}$ would be opposite between $\Lambda$ and $\bar{\Lambda}$, 
and this property might be employed to distinguish the signal from background contribution.

\begin{acknowledgements}
%
We are grateful to
Koichi Hattori,
Xionghong He,
Shu Lin,
Hao Qiu,
Chun Shen,
Pu Shi,
Subhash Singha,
Xin-Li Sheng,
Shusu Shi,
Shuzhe Shi,
Qun Wang,
Nu Xu
Naoki Yamamoto,
Ho-Ung Yee,
Jie Zhao
for helpful conversations. We in particular thank Longgang Pang for drawing us the attention to Ref.~\cite{Wu:2019eyi}, which motivates this work. 
This work was supported by the Strategic Priority Research Program of Chinese Academy of Sciences, Grant No. XDB34000000. 
\end{acknowledgements}

%
%

\appendix

\bibliography{refs}

\end{document}